\documentclass[aps,prx, superscriptaddress, twocolumn, tightenlines, longbibliography, showpacs, floatfix, pdftex]{revtex4-1}

\pdfoutput=1

\usepackage{amsmath, amsfonts, amssymb}
\usepackage{bm,bbm}
\usepackage{mathtools}

\usepackage{afterpackage}
\usepackage{hyperref}
\hypersetup{
    colorlinks,
    linkcolor=blue,
    citecolor=blue,
    filecolor=blue,
    urlcolor=blue
}
\usepackage{multirow}
\usepackage{tikz} 
\usepackage[normalem]{ulem}
\usepackage{wasysym}

\usepackage{float}
\usepackage{graphicx}
\usepackage[caption=false]{subfig} 
\usepackage{array, booktabs}
	\newcolumntype{P}[1]{>{\centering\arraybackslash}p{#1}} 

\newcommand{\corr}[1]{\langle #1\rangle}

\newcommand{\mb}[1]{\mathbf{#1}}

\newcommand{\bds}[1]{\boldsymbol{#1}}

\newcommand{\abs}[1]{\vert #1\vert}

\newcommand{\commentblock}[1]{} 

\newcolumntype{C}{>{$}c<{$}}
\AtBeginDocument{
\heavyrulewidth=.08em
\lightrulewidth=.05em
\cmidrulewidth=.03em
\belowrulesep=.65ex
\belowbottomsep=0pt
\aboverulesep=.4ex
\abovetopsep=0pt
\cmidrulesep=\doublerulesep
\cmidrulekern=.5em
\defaultaddspace=.5em
}

\begin{document}

\title{Machine-Learned Phase Diagrams of Generalized Kitaev Honeycomb Magnets}

\date{\today}

\author{Nihal Rao}
\affiliation{Arnold Sommerfeld Center for Theoretical Physics, University of Munich, Theresienstr. 37, 80333 M\"unchen, Germany}
\affiliation{Munich Center for Quantum Science and Technology (MCQST), Schellingstr. 4, 80799 M\"unchen, Germany}

\author{Ke Liu}
\email{ke.liu@lmu.de}
\affiliation{Arnold Sommerfeld Center for Theoretical Physics, University of Munich, Theresienstr. 37, 80333 M\"unchen, Germany}
\affiliation{Munich Center for Quantum Science and Technology (MCQST), Schellingstr. 4, 80799 M\"unchen, Germany}

\author{Marc Machaczek}
\affiliation{Arnold Sommerfeld Center for Theoretical Physics, University of Munich, Theresienstr. 37, 80333 M\"unchen, Germany}
\affiliation{Munich Center for Quantum Science and Technology (MCQST), Schellingstr. 4, 80799 M\"unchen, Germany}

\author{Lode Pollet}
\affiliation{Arnold Sommerfeld Center for Theoretical Physics, University of Munich, Theresienstr. 37, 80333 M\"unchen, Germany}
\affiliation{Munich Center for Quantum Science and Technology (MCQST), Schellingstr. 4, 80799 M\"unchen, Germany}
\affiliation{Wilczek Quantum Center, School of Physics and Astronomy, Shanghai Jiao Tong University, Shanghai 200240, China}

\begin{abstract}
We use a recently developed interpretable and unsupervised machine-learning method, the tensorial kernel support vector machine (TK-SVM), to investigate the low-temperature classical phase diagram of a generalized Heisenberg-Kitaev-$\Gamma$ ($J$-$K$-$\Gamma$) model on a honeycomb lattice.
Aside from reproducing phases reported by previous quantum and classical studies, our machine finds a hitherto missed nested zigzag-stripy order and establishes the robustness of a recently identified modulated $S_3 \times Z_3$ phase, which emerges through the competition between the Kitaev and $\Gamma$ spin liquids, against Heisenberg interactions. 
The results imply that, in the restricted parameter space spanned by the three primary exchange interactions---$J$, $K$, and $\Gamma$, the representative Kitaev material $\alpha$-${\rm RuCl}_3$ lies close to the boundaries of several phases, including a simple ferromagnet, the unconventional $S_3 \times Z_3$ and nested zigzag-stripy magnets.
A zigzag order is stabilized by a finite $\Gamma^{\prime}$ and/or $J_3$ term, whereas the four magnetic orders may compete in particular if $\Gamma^{\prime}$ is anti-ferromagnetic.
\end{abstract}

\maketitle

\section{Introduction}\label{sec:intro}
Machine learning (ML) is quickly developing into a powerful tool in modern day physics research~\cite{Carleo19, Mehta19}.
Successful applications in condensed-matter physics can be found in, for example, detecting phases and phase transitions~\cite{Wang16, Ponte17, Carrasquilla17, Chen20}, representing and solving quantum wave functions~\cite{Carleo17, Carrasquilla20, Deng17, Hermann20, Pfau20, Vieijra20}, analyzing experiments~\cite{Nussinov16, Zhang19, Bohrdt19}, searching new materials~\cite{Schmidt19}, and designing algorithms~\cite{Liao19, LiuJW17}. 
Recent developments of ML in strongly correlated condensed matter physics are moving beyond benchmarking, and the ultimate goal is to provide  toolboxes to tackle hard and open problems.

The Kitaev materials~\cite{Takagi19, Janssen19, Winter17b} are prime candidates for a challenging application of ML,  hosting various disordered and unconventionally ordered phases.
Experimentally, the bond-dependent anisotropic interactions of the Kitaev honeycomb model~\cite{Kitaev06} are realized through electron correlations and spin-orbit coupling~\cite{Jackeli09, Chaloupka10}.
Representative compounds include $4d$ and $5d$ transition-metal-based Mott insulators ${\rm A}_2 {\rm IrO}_3$ ($\text{A} = \text{Na, Li, K}$) and $\alpha$-${\rm RuCl}_3$~\cite{Winter17b, Banerjee16, Banerjee17, Ran17, Yadav16, Yadav19, Chaloupka15}.
In particular, the latter material has been proposed to host a field-induced quantum spin liquid as evidenced by the half-quantized thermal Hall effect under external magnetic field~\cite{Kasahara18, Yokoi20}, while spectroscopic~\cite{Ponomaryov20, Sahasrabudhe20, Maksimov20} and thermodynamic~\cite{Bachus20, Bachus21} measurements indicate a topologically trivial partially-polarized phase.
More recently, the cobaltate systems ${\rm Na}_3{\rm Co}_2 {\rm SbO}_6$ and  ${\rm Na}_2{\rm Co}_2 {\rm TeO}_6$~\cite{Viciu07, LiuHM18, LiuHM20, Songvilay20} and Cr-based  pseudospin-$3/2$ systems ${\rm CrSiTe}_3$ and ${\rm CrGeTe}_3$~\cite{Xu20} were added to this family.

In the ideal case, one expects to find a compound that faithfully exhibits the physics of the Kitaev model.
However, non-Kitaev terms, such as the Heisenberg exchange and the symmetric off-diagonal $\Gamma$ exchange, are permitted by the underlying cubic symmetry and ubiquitously exist in real Kitaev materials~\cite{Rau14, Rau16}.
In addition, longer-range interactions and structural distortions can lead to further hopping channels~\cite{Rau14b, Rusnacko19, Maksimov20}. 
These additional terms enrich the Kitaev physics~\cite{Wang19, Gohlke18, Gohlke20, Chern20, Lee20, Jiang19, Gordon19, Rusnacko19, Lampen18, Liu21} but also pose a significant challenge to the analysis because of the large parameter space and the emergence of complicated structures.
Therefore, tools that can efficiently detect patterns and important information in data and construct the associated phase diagrams are called for.

In this work, we use our recently developed tensorial-kernel support vector machine (TK-SVM)~\cite{Greitemann19, Liu19, Greitemann19b} to investigate the phase diagram of a generalized Heisenberg-Kitaev-$\Gamma$ model on a honeycomb lattice.
This method is \emph{interpretable} and \emph{unsupervised}, equipped with a tensorial kernel and graph partitioning.
The tensorial kernel detects both linear and high-order correlations, and the results can systematically be interpreted as meaningful physical quantities, such as order parameters~\cite{Greitemann19} and emergent local constraints~\cite{Greitemann19b}. 
Moreover, in virtue of the graph partitioning module, phase diagrams can be constructed without supervision and prior knowledge.
 
In our previous investigation of the Kitaev magnets we applied TK-SVM to the classical $K$-$\Gamma$ model subject to a magnetic field~\cite{Liu21}.
There, our machine learned a rich global phase diagram, revealing, among others, two novel modulated $S_3 \times Z_3$ phases, which originate from the competition between the Kitaev and $\Gamma$ spin liquids.
This work explores the low-temperature classical phase diagram of the generic Heisenberg-Kitaev-$\Gamma$ ($J$-$K$-$\Gamma$) model as well as the effect of the $\Gamma^{\prime}$ and third nearest-neighbor Heisenberg ($J_{3}$) terms, which are sub-leading exchange terms commonly encountered in the class of Kitaev materials.
From our findings it follows that in the parameter space spanned by $J$,$K$, and $\Gamma$, the representative Kitaev material $\alpha$-${\rm RuCl}_3$ lies close to several competing phases, including a hitherto missed nested zigzag-stripy magnet, a previously identified $S_3 \times Z_3$ magnet, a ferromagnet, and a possibly correlated paramagnet (Section~\ref{sec:jkgm}). 
Zigzag order can be stabilized by including a small $\Gamma^{\prime}$ and/or anti-ferromagnetic $J_3$ term.
However, if the $\Gamma^{\prime}$ is also antiferromagnetic, this material resides in a region where these four magnetic orders strongly compete (Section~\ref{sec:j3gmp}).
Our results constitute therefore one of the earliest examples of ML going beyond the state of the art in strongly correlated condensed matter physics.

This paper is organized as follows. In Section~\ref{sec:model}, we define the generalized Heisenberg-Kitaev-$\Gamma$ model and specify the interested parameter regions.
The machine-learned $J$-$K$-$\Gamma$ phase diagrams in the absence and presence of additional $J_3$ and $\Gamma^{\prime}$ terms are discussed and validated in Section~\ref{sec:jkgm} and Section~\ref{sec:j3gmp}, respectively.
Section~\ref{sec:materials} discusses the implications of our results for representative Kitaev compounds.
In Section~\ref{sec:summary} we  conclude and provide an outlook. 
In addition, a brief summary of TK-SVM and details about the training procedure and Monte Carlo simulations are provided in Appendices.

\section{Honeycomb $J$-$K$-$\Gamma$-$\Gamma^{\prime}$-$J_3$ Model}\label{sec:model}

\begin{figure}[t]
  \centering
  \includegraphics[width=0.3\textwidth]{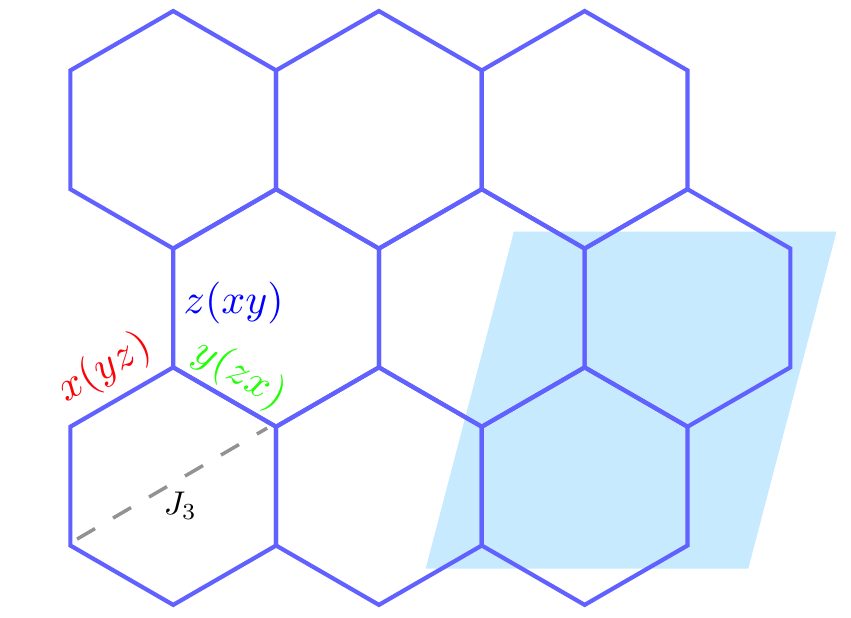}
  \caption{A honeycomb lattice with anisotropic bonds $\gamma(\alpha\beta)$. The shaded region marks a symmetric cluster of $m\times m$ unit cells. A lattice with linear size $L$ is then partitioned into $(\frac{L}{m})^2$ such clusters. Here, $m = 2$ is shown for example. Larger clusters with $m = 4, 6, 12$ are considered in training TK-SVMs.}
  \label{fig:lattice}
\end{figure}

We study the generalized Heisenberg-Kitaev-$\Gamma$ model on a honeycomb lattice
\begin{align}\label{eq:model}
 H  & = H_{JK\Gamma} + H_{\Gamma^{\prime}} + H_{J_3}, \nonumber \\
 	& = \sum_{\corr{ij}_\gamma} \mb{S}_i \cdot \hat{\mathcal{J}}_\gamma \mb{S}_j+ \sum_{(ij)} J_3 \mb{S}_i \cdot \mb{S}_j,
\end{align}
where
\begin{align}
 & H_{\scalebox{0.618}{$JK\Gamma$}} \! = \! \sum_{\corr{ij}_\gamma} \big[ 
J \mb{S}_i \cdot \mb{S}_j \! + \!K S_i^\gamma S_j^\gamma \! + \! \Gamma (S_i^\alpha S_j^\beta + S_i^\beta S_j^\alpha) \big], \label{eq:jkgm} \\
 & H_{\Gamma^{\prime}} = \sum_{\corr{ij}_\gamma} \big[ \Gamma^{\prime} (S_i^\gamma S_j^\alpha + S_i^\gamma S_j^\beta + S_i^\alpha S_j^\gamma + S_i^\beta S_j^\gamma) \big].
\end{align}
Here, $\gamma$ labels the three distinct nearest-neighbor (NN) bonds $\corr{ij}$ with mutually exclusive $\alpha, \beta, \gamma \in \{x, y, z \}$ as illustrated in Figure~\ref{fig:lattice};
 $\hat{\mathcal{J}}_\gamma$ is a $3\times 3$ matrix comprising all exchanges on a NN bond $\corr{ij}_\gamma$, and
$(ij)$ denotes the third NN bonds with a Heisenberg interaction $J_3$. 

The Heisenberg-Kitaev-$\Gamma$ Hamiltonian Eq.~\eqref{eq:jkgm} comprises generic NN exchanges allowed by the cubic symmetry~\cite{Jackeli09, Rau14}.
Although the Kitaev ($K$) term is of prime interest for realizing quantum Kitaev spin liquids,  the Heisenberg ($J$) and the symmetric off-diagonal ($\Gamma$) exchanges ubiquitously exist and  play a key role in the physics of realistic materials.    
The $\Gamma^{\prime}$ term is a secondary symmetric off-diagonal interaction and originates from a trigonal distortion of the octahedral environment of magnetic ions. 
A negative (positive) $\Gamma^{\prime}$ corresponds to trigonal compression (expansion) of the edge-sharing oxygen or chlorine octahedra~\cite{Rau14b} while the inclusion of the  $J_3$ term reflects the extension of $d$-electron wave functions.
Although second nearest-neighbor exchanges are also possible, the third-neighbor exchanges are found to be more significant in representative Kitaev magnets, including the intensely studied compounds ${\rm Na}_2 {\rm IrO}_3$, $\alpha$-${\rm Li}_2 {\rm IrO}_3$, $\alpha$-${\rm RuCl}_3$ and the more recently (re-)characterized cobalt-based compounds ${\rm Na}_3{\rm Co}_2 {\rm SbO}_6$ and  ${\rm Na}_2{\rm Co}_2 {\rm TeO}_6$~\cite{Viciu07, Songvilay20}.
Aside from the potential microscopic origin, the $\Gamma^{\prime}$ and $J_3$ exchange terms are often introduced phenomenologically to stabilize magnetic orders observed in experiments~\cite{Banerjee17, Rusnacko19, Laurell20, Maksimov20}, in particular the zigzag-type orders found in many two-dimensional Kitaev materials~\cite{Takagi19}.

It is commonly expected that the primary physics in a Kitaev material is governed by the interactions in the $H_{JK\Gamma}$ model, whose phase diagram for fixed $\Gamma^{\prime}$ and $J_3$ is the topic of the present work.
Moreover, motivated by the microscopic models proposed for $\alpha$-${\rm RuCl}_3$~\cite{Maksimov20, Laurell20},  ${\rm Na}_3{\rm Co}_2 {\rm SbO}_6$ and  ${\rm Na}_2{\rm Co}_2 {\rm TeO}_6$~\cite{Songvilay20} (cf. Section~\ref{sec:j3gmp}), we focus on the parameter space with $K < 0$, $\Gamma > 0$ and a moderate range of ferromagnetic Heisenberg ($J$) exchange terms.

\begin{figure}[t]
  \centering
  \includegraphics[width=0.47\textwidth]{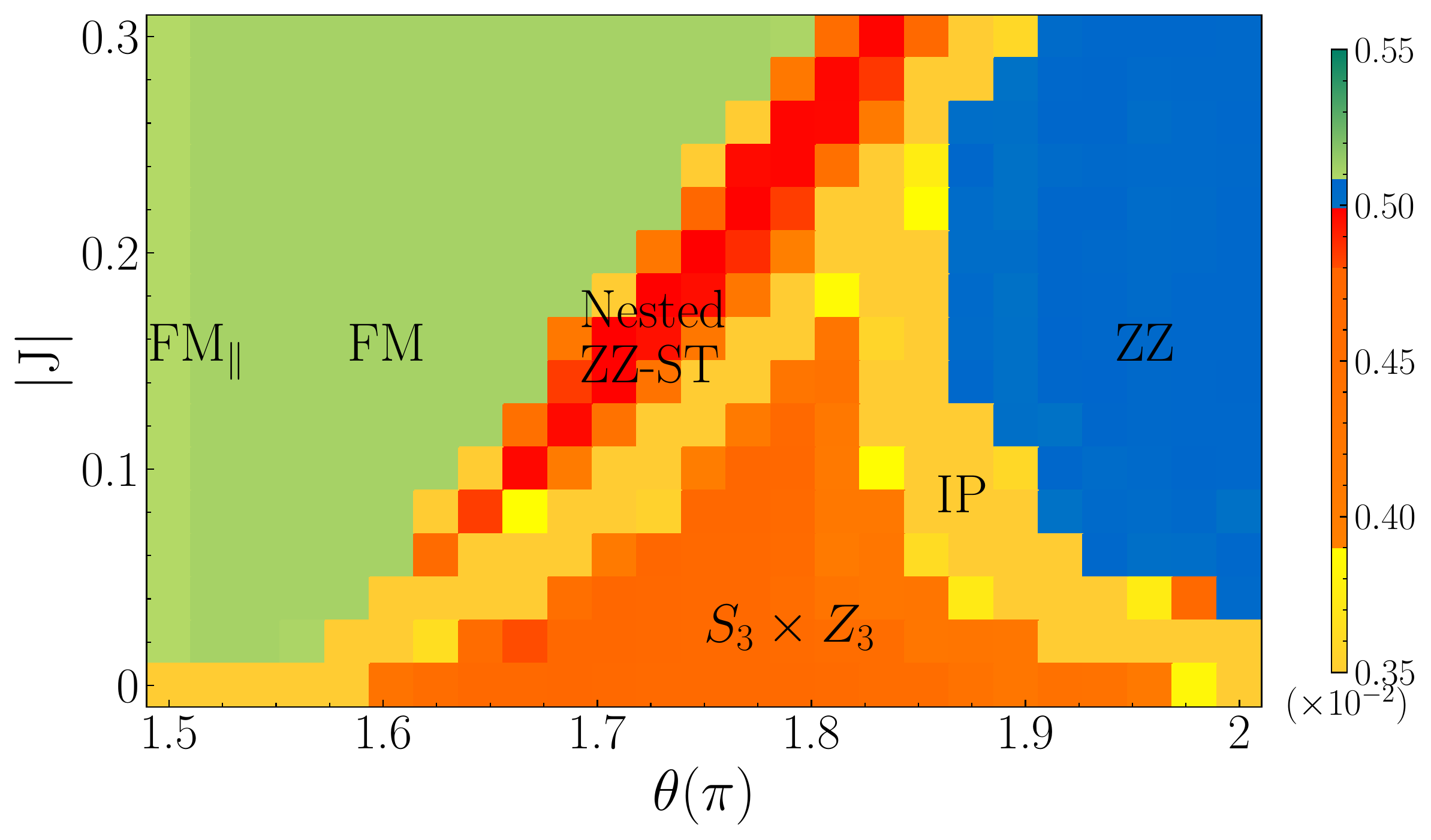}
  \caption{ Machine-learned $J$-$K$-$\Gamma$ phase diagram for parameters $J < 0, \ K = \sin{\theta} < 0, \ \Gamma = \cos{\theta} >0$, at $T = 10^{-3}$. Interactions and temperature are in units of $\sqrt{K^2 + \Gamma^2}$.
   Each pixel represents a $(\theta, J)$ point with $\Delta \theta = \frac{1}{48} \pi$ and $\Delta J = 0.02$; same for the phase diagrams below.
        A rank-$1$ TK-SVM with symmetric cluster of $12\times 12$ lattice cells is used. 
  	The color represents the Fiedler entry value (FEV) for the corresponding $(\theta, J)$ point, and the choice of the color bar is guided by the histogram of FEVs (Appendix~\ref{app:learning}).
    Parameter points in the same phase have the same or very close values. The blurry regions indicate phase boundaries and crossovers.
    The Kitaev and $\Gamma$ spin liquids reside at the corner of $(\theta, J) = (\frac{3}{2} \pi, 0)$ and $(2\pi, 0)$, respectively, which are not distinguished from disordered IP regime as the rank-$1$ TK-SVM detects magnetic orders.
  	FM: ferromagnetic, where $\text{FM}_\parallel$ indicates easy-axis states; Nested ZZ-ST: nested zigzag-stripy; IP: incommensurate or (correlated) paramagnetic.}
  \label{fig:pd_0_0}
\end{figure}

Specifically, we parametrize the Kitaev and $\Gamma$ interactions as $K = \sin{\theta}, \ \Gamma = \cos{\theta}$, scan over $\theta \in [\frac{3}{2}\pi, 2\pi]$ and restrict the Heisenberg interaction to  $ J \in [-0.3, 0]$.
We investigate slices of experimental relevance with $J_3 = 0, 0.1$ and $\Gamma^{\prime} = 0, \pm 0.1$.
In particular, considering a ferromagnetic as well as an anti-ferromagnetic $\Gamma^{\prime}$ covers both cases of its disputed sign in $\alpha$-${\rm RuCl}_3$.
Most of the previous studies considered a negative $\Gamma^{\prime}$ (trigonal compression)~\cite{Kim16, Winter16, Eichstaedt19, Sears20}.
However, a recent work Ref.~\onlinecite{Maksimov20} advocates a positive $\Gamma^{\prime}$ (trigonal expansion) by stressing the  electron-spin-resonance (ESR)~\cite{Ponomaryov17} and terahertz (THz)~\cite{Sahasrabudhe20} experiments, and its critical magnetic fields~\cite{Cao16, Lampen18, Winter18}.

We treat spins as  $O(3)$ vectors to gain training data for large system sizes, corresponding to the classical large-$S$ limit.
We employ parallel-tempering Monte Carlo simulations with a heat bath algorithm and over relaxation to generate spin configurations and simulate large system sizes up to $10,386$ spins ($72\times 72$ honeycomb unit cells), to accommodate potential competing orders.
During the training procedure, $400$ $(\theta, J)$ points are simulated for each fixed $\Gamma^{\prime}$ and $J_3$ slice, and in total $2,400$ points are simulated.
Training samples are collected at low temperature $T = 10^{-3} \sqrt{K^2 + \Gamma^2}$.
Classification of these parameter points unravels the topology of the $J$-$K$-$\Gamma$ phase diagram for each of the six $\Gamma^{\prime}$ and $J_3$ combinations.
Thereafter, we extract the physical order parameter of each phase from the learned TK-SVM decision functions.
These order parameters are then measured in new simulations down to the temperature $T = 10^{-4} \sqrt{K^2 + \Gamma^2}$, in the most frustrated parameter regimes and passing through different phases and phase boundaries.
The nature of the phases as well as the topologies of the machine-learned phase diagrams are thereby verified.
See Appendix~\ref{app:learning} for the setup of the sampling and training and Appendix~\ref{app:simulation} for details of the Monte Carlo simulations.

It turns out that the phase diagrams of the investigated parameter regions are dominated by various magnetic orders.
This indicates that the classical phase diagrams may qualitatively, or even semi-quantitatively, reflect those of finite spin-$S$ values.
Indeed, we successfully reproduce the ferromagnetic, zigzag and $120^{\circ}$ orders previously observed in quantum and classical analysis~\cite{Rau14, Rusnacko19, Maksimov20} and in addition find more phases.

\begin{figure}[t]
  \centering
  \includegraphics[width=0.3\textwidth]{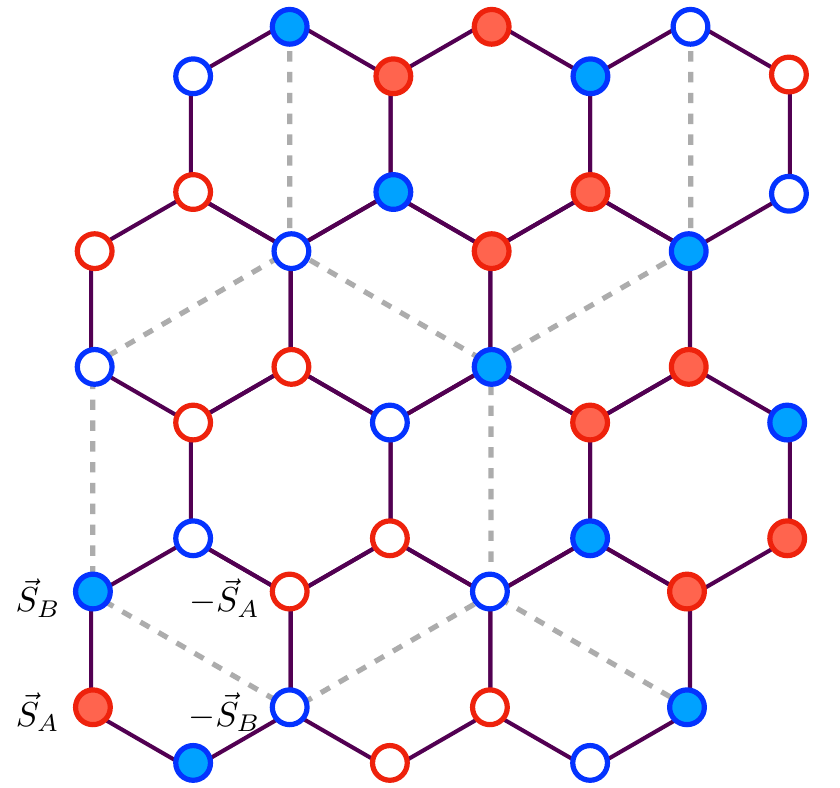}
  \caption{A ground-state configuration of the {\it nested} zigzag-stripy order. The red ($A$) and blue ($B$) colors label two inequivalent reference spins, $\vec{S}_A \neq \vec{S}_B$. The filled ($+$) and empty ($-$) cycles indicate the sign of a spin. Here the $A$-spins ($B$-spins) form zigzag (stripy) structures on a honeycomb lattice with a doubled lattice spacing. The dashed lines are a guide to the eye.}
  \label{fig:nested_pattern}
\end{figure}

\begin{figure}[t]
 \centering
 \includegraphics[width=0.4\textwidth]{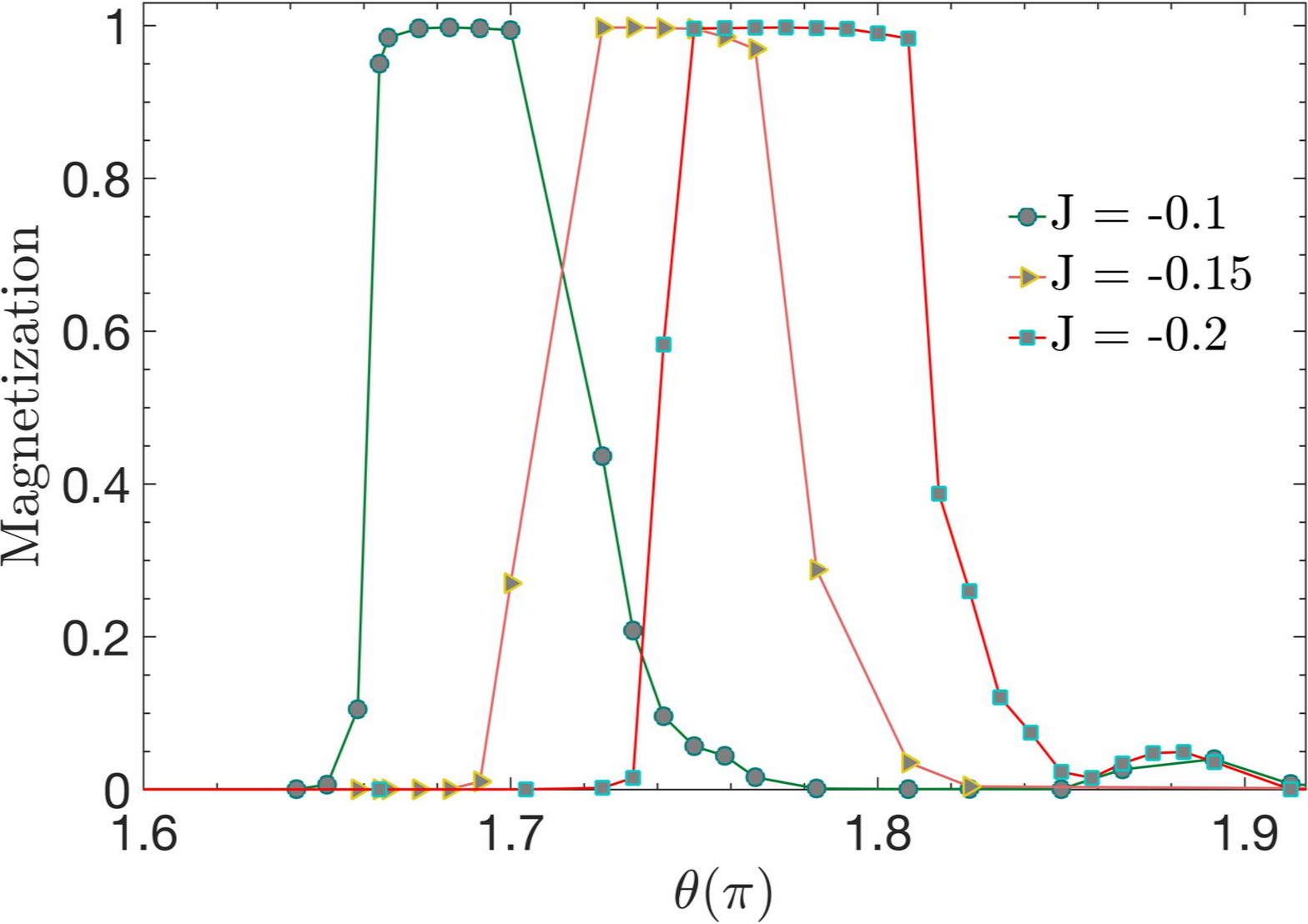}
 \caption{Monte Carlo measurements of the nested zigzag-stripy magnetization at different $J$'s, with $\Gamma^{\prime} = J_3 = 0, \ T = 10^{-4}$.
 		Consistent with the phase diagram Figure~\ref{fig:pd_0_0} learned at $T = 10^{-3}$, the nested zigzag-stripy  order is preferred by larger $\abs{J}$ and $\Gamma$.}\label{fig:mag_nest}
 \end{figure}

\begin{figure*}
  \centering
  \includegraphics[width=0.75\textwidth]{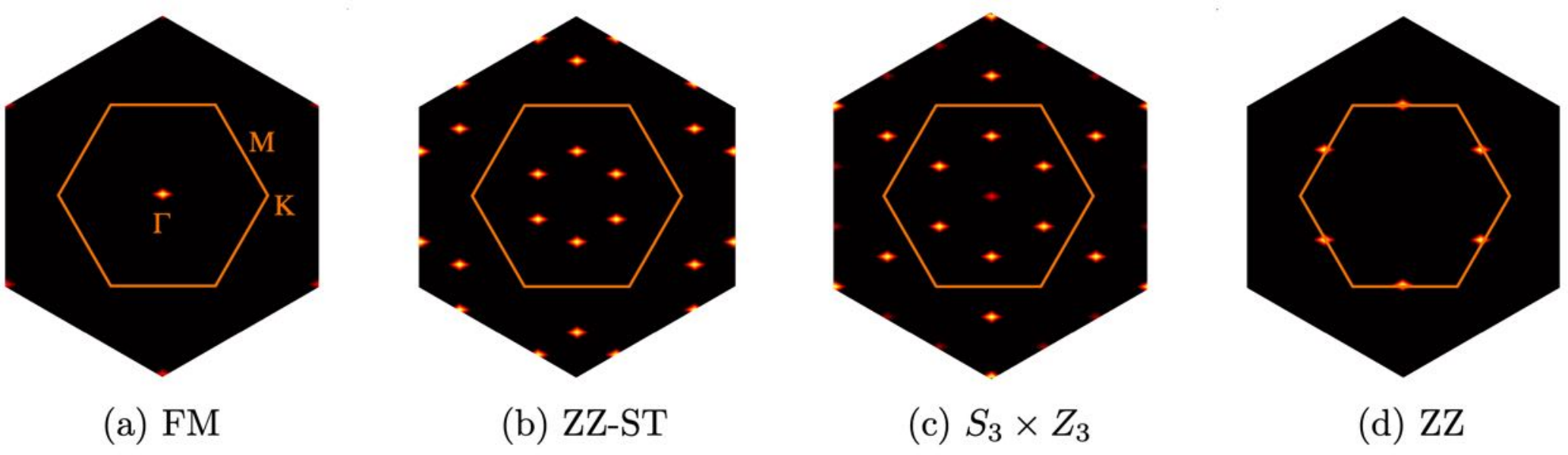}
  \caption{Evolution of the spin structure factor $S (\mb{q})$. The inner (outer) area denotes the first (second) honeycomb Brillouin zone; high symmetries points are indicated. Here
   $S (\mb{q}) =  \big\langle \frac{1}{2L^2} \sum_{ij} \mb{S}_i \cdot \mb{S}_j \, e^{i \mb{q} \cdot (\mb{r}_i - \mb{r}_j)} \big\rangle$
   is measured at $J=-0.1$ and $T = 10^{-3}$. Upon increasing $\Gamma$, the magnetic Bragg peaks pass by the $\mb{\Gamma}$ (FM), $\frac{1}{2} \mb{M}$ (nested ZZ-ST), $\frac{2}{3} \mb{M}$ (modulated $S_3 \times Z_3$) and $\mb{M}$ (zigzag) points. The length of the wave factors are stable within each phase.}
  \label{fig:SSF}
\end{figure*}

\begin{figure*}
 	 \centering
 	 \includegraphics[width=0.95\textwidth]{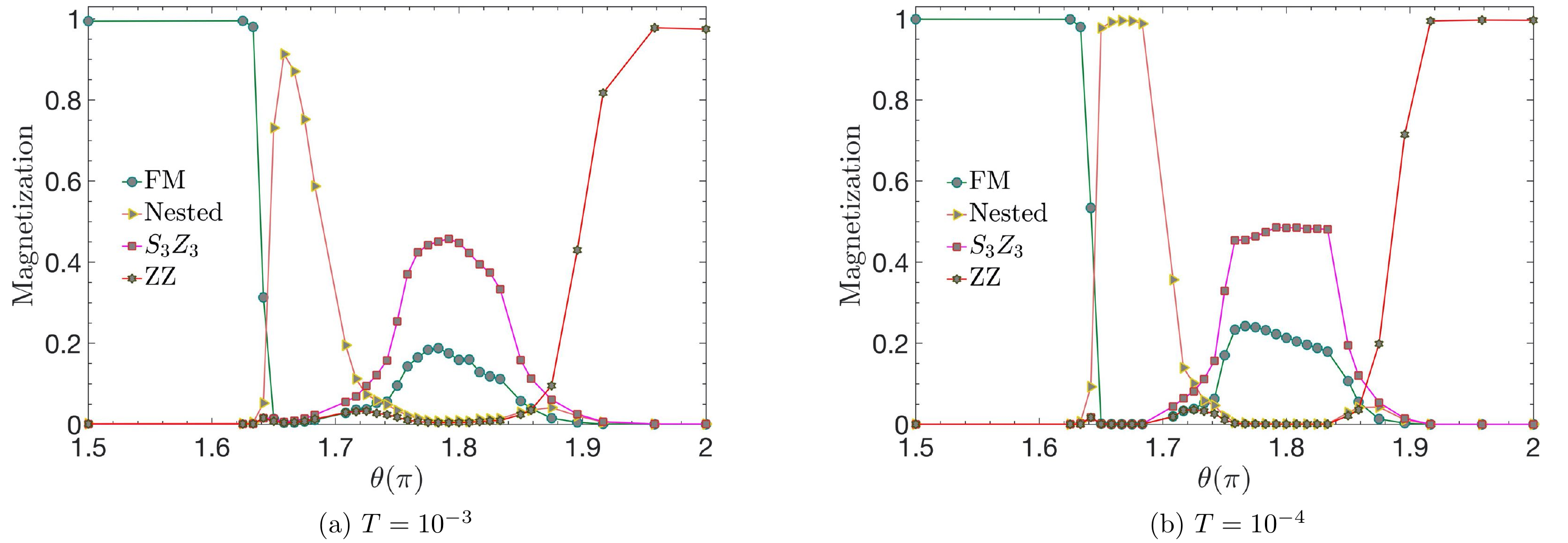}
 \caption{Monte Carlo measurements of the magnetization at fixed $J = -0.1, \ \Gamma^{\prime} = J_3 = 0$. Results for $T = 10^{-3}$ and $T = 10^{-4}$ are compared. The magnetic order of each phase in Figure~\ref{fig:pd_0_0} is  confirmed.
  The broad IP regions are narrower at very low temperature $T = 10^{-4}$ but remain quite sizable, indicating that these regions are highly frustrated and the orders fragile.}\label{fig:mags}
 \end{figure*}

\section{$J$-$K$-$\Gamma$ Phase Diagram}\label{sec:jkgm}
We focus in this section on the machine-learned phase diagram for the pure Heisenberg-Kitaev-$\Gamma$ model and save the discussion on the effects of the $\Gamma^{\prime}$ and $J_3$ terms for Section~\ref{sec:j3gmp}.

The $J$-$K$-$\Gamma$ phase diagram has previously been explored by several authors; see, for example, Refs.~\onlinecite{Rau14, Wang19, Rusnacko19, Chaloupka15}.
In the parameter regions with dominating Heisenberg and Kitaev exchanges, different methods give consistent results.
The ferromagnetic, zigzag, anti-ferromagnetic, and stripy orders in the $J$-$K$ phase diagram~\cite{Chaloupka10, Janssen16} extend to regions of finite $\Gamma$~\cite{Rau14, Rusnacko19}.
The physics is however more subtle when the system is governed by competing Kitaev and $\Gamma$ interactions.
In the parameter regime with $K < 0$, $\Gamma \sim \abs{K}$ and a small but finite ferromagnetic $J$ term, believed to be relevant for $\alpha$-${\rm RuCl}_3$, a previous study based on a Luttinger-Tisza analysis suggests a zigzag order~\cite{Rau14}.
However, this order is not confirmed by the $24$-site exact diagonalization (ED) carried out in the same work, and a more recent study~\cite{Rusnacko19} equipped with $32$-site ED and cluster mean-field calculations shows that the physics depends on the size and shape of clusters.

Our machine finds that the phase diagram in the above parameter regime is quite rich, as shown in Figure~\ref{fig:pd_0_0}.
In addition to reproduce the ferromagnetic and zigzag phase in the large $K$ and $\Gamma$ regions~\cite{Rau14, Wang19, Rusnacko19} under a finite $J$, our machine also identifies a novel {\it nested} zigzag-stripy (ZZ-ST) phase and shows the extension of the $S_3 \times Z_3$ phase. 
The $S_3 \times Z_3$ phase results from the competition between the Kitaev and $\Gamma$ spin liquids and features a spin-orbit entangled modulation, with magnetic Bragg peaks at $\frac{2}{3}\mathbf{M}$ points~\cite{Liu21}.

\begin{figure}[t]
 \centering
 \includegraphics[width=0.5\textwidth]{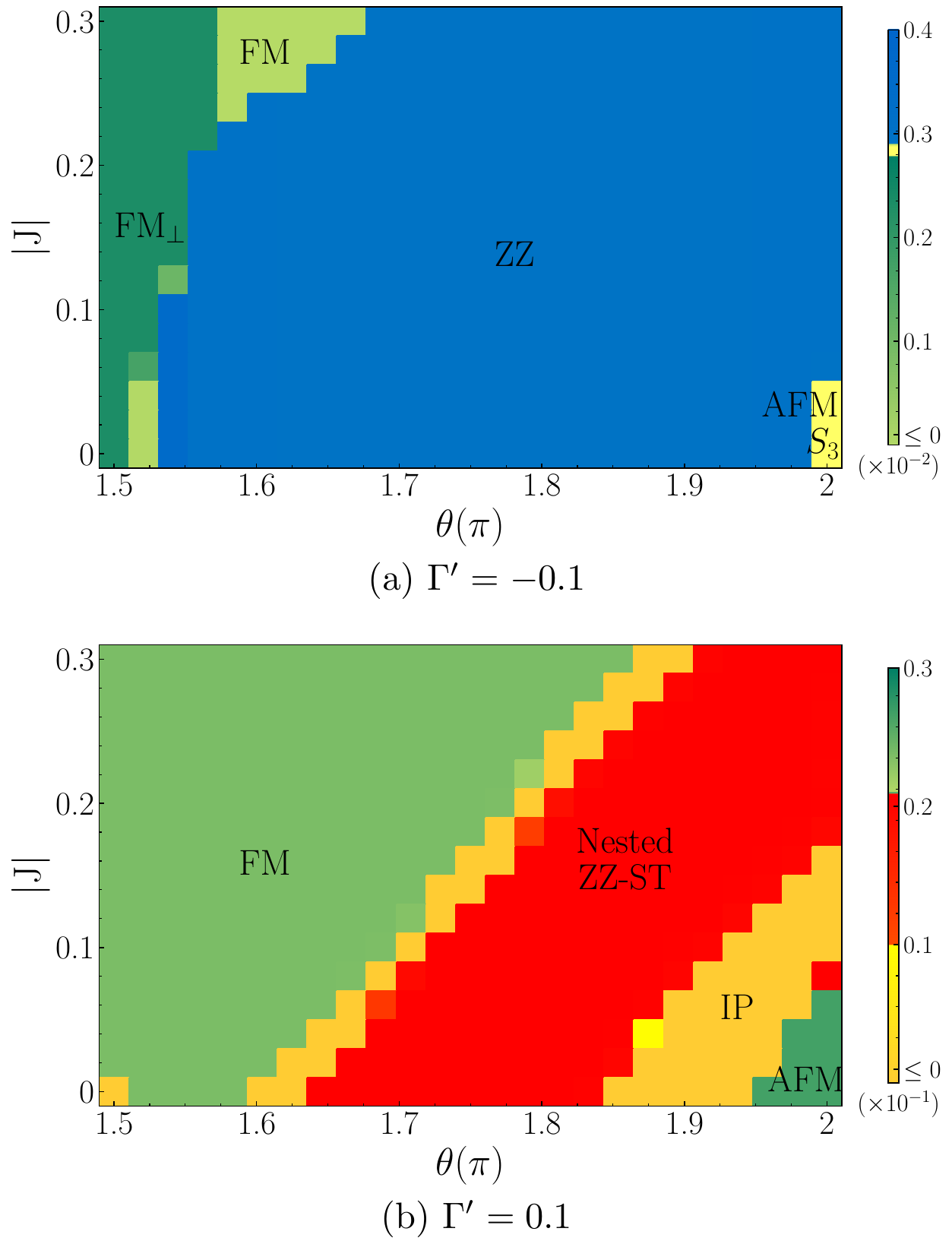}
 \caption{Machine-learned $J$-$K$-$\Gamma$ phase diagram, with $J_3 = 0, \ \Gamma^{\prime} = \pm 0.1$ at $T = 10^{-3}$. A zigzag phase prevails over the phase diagram when a ferromagnetic $\Gamma^{\prime} = -0.1$ is considered, while an anti-ferromagnetic $S_3$ order is stabilized in the large $\Gamma$ limit. All orders in (a) are unfrustrated.
 By contrast, in the case of an anti-ferromagnetic $\Gamma^{\prime}$ (b), the nested zigzag-stripy phase expands significantly, and there remains a highly frustrated IP region   at larger $\Gamma$ values.
 Panels (a) and (b) are learned with a symmetric $6 \times 6$  and $4 \times 4$ cluster, respectively.
The resolution of $(\theta, J)$ points is same as in Fig.~\ref{fig:pd_0_0}, namely, $\Delta \theta = \frac{1}{48} \pi$ and $\Delta J = 0.02$. 
 The corresponding Fiedler entries are color plotted in arbitrary unit, and their histograms are provided in Appendix~\ref{app:learning}.
 }
 \label{fig:pd_gmp}
 \end{figure}

The nested ZZ-ST order has not been reported in previous studies to the best of our knowledge.
In this phase, whose representative ground-state configuration is illustrated in Figure~\ref{fig:nested_pattern}, spins can be divided into two groups, $\{\vec{S}_{A}, \vec{S}_{B}\}$.
One set of spins, e.g., the $A$-spins in Figure~\ref{fig:nested_pattern}, form regular zigzag patterns with a doubled lattice constant while the other set of spins ($B$-spins) form stripy patterns, intricately nested with the zigzag pattern of the $A$-spins.
This nesting of orders enlarges the ground-state manifold: The global three-fold rotation $(C_{3})$ and spin-inversion symmetry $(S \rightarrow -S)$ of the (generalized) $J$-$K$-$\Gamma$ model trivially allows six ground states.
This degeneracy is further doubled as the two sets of spins can be swapped, leading to twelve distinct ground states, which have all been observed in our Monte Carlo simulations.
In addition, the robustness of the order is also confirmed in Figure~\ref{fig:mag_nest} by scanning $\theta$ at different $J$ values.

The formation of the $S_3 \times Z_3$ and the nested ZZ-ST orders leads to an interesting evolution in spin structure factors (SSFs).
As shown in Figure~\ref{fig:SSF} for a fixed \mbox{$J = -0.1$}, in the ferromagnetic phase at small $\Gamma$, the magnetic Bragg peak develops at the $\mb{\Gamma}$ point of the honeycomb Brillouin zone.
Increasing the $\Gamma$ coupling results in the magnetic Bragg peaks moving outwards to the $\frac{1}{2} \mb{M}$, $\frac{2}{3} \mb{M}$ and $\mb{M}$ points, as the system passes the nested ZZ-ST, $S_3 \times Z_3$, and zigzag orders, respectively.

These phases are nonetheless separated by broad crossover areas, marked as incommensurate or paramagnetic (IP) regimes, where our machine does not detect any clear magnetic ordering down to the temperature $T = 10^{-3}$.
Explicit measurements of the learned order parameters at a lower temperature $T = 10^{-4}$ further show all magnetic moments are indeed remarkably fragile, as plotted in Figure~\ref{fig:mags} with a fixed $J = -0.1$.
Although with training data from a finite-size system and finite temperature, we cannot exclude lattice incommensuration and long-range orders at $T \rightarrow 0$ in these areas, our system size is considerably large and the absence of stable magnetic orders at such low temperatures is quite notable.
One can expect quantum fluctuations will be enhanced in these areas as classical orders are suppressed, potentially hosting quantum paramagnets or spin liquids for finite spin-$S$ systems.
The Kitaev and $\Gamma$ spin liquids are not distinguished from the disordered IP regimes in the phase diagram Figure~\ref{fig:pd_0_0} as the rank-$1$ TK-SVM detects only magnetic correlations.
However, as we studied in Ref.~\onlinecite{Liu21} for the $K$-$\Gamma$ model, while a classical $\Gamma$SL are less robust against competing interactions, a classical KSL can thermally extend to a finite area.

\begin{figure}[t]
 \centering
 \includegraphics[width=0.4\textwidth]{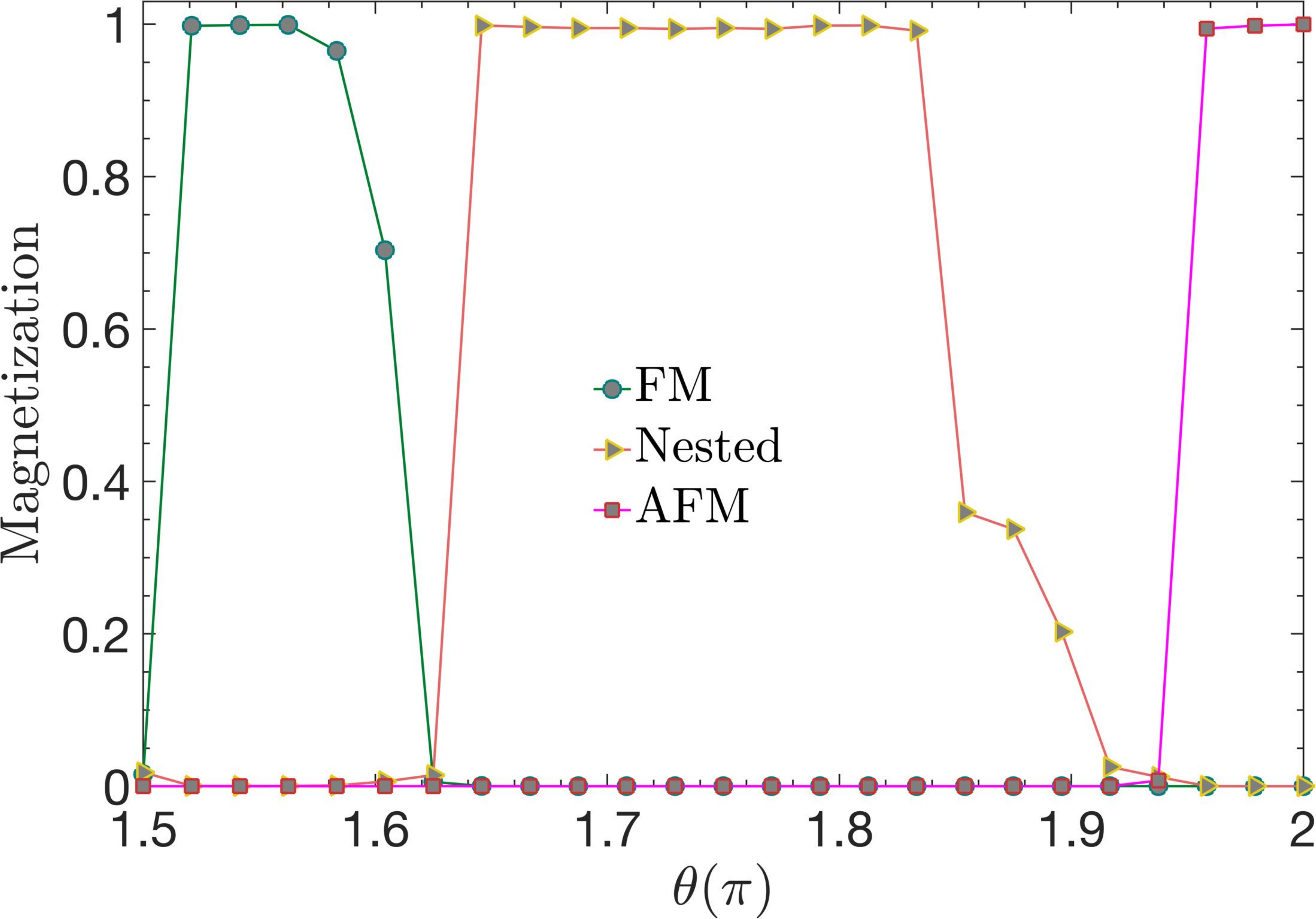}
 \caption{Monte Carlo measurements of the magnetization with fixed $J = 0, \ \Gamma^{\prime} = 0.1, \ J_3 = 0$, at $T = 10^{-4}$. The wide window between the nested zigzag-stripy and anti-ferromagnetic orders corresponds to the IP regime in the phase diagram Figure~\ref{fig:pd_gmp} (b).}
 \label{fig:mags_gmp}
 \end{figure}

\begin{figure*}
\includegraphics[width=1\textwidth]{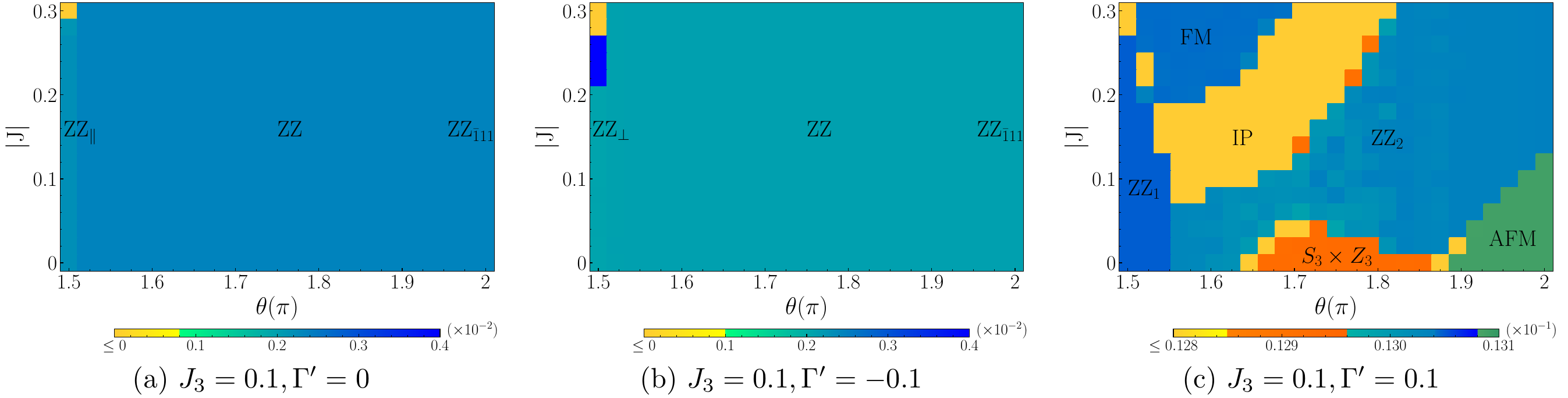}
\caption{Machine-learned phase diagrams with $J_3 = 0.1, \, \Gamma^{\prime} = 0, \, \pm 0.1$ at $T = 10^{-3}$.
	The $J_3$ term universally prefers zigzag states. In cases of a vanishing and negative $\Gamma^{\prime}$, zigzag phases dominate the phase diagram, except for a narrow IP window at the $\Gamma = 0$ limit with $J_3/J \sim -\frac{1}{3}$, which may be a remnant of a spiral order in the $J_1$-$J_3$ honeycomb model.
	Special zigzag states are marked in (a) and (b). With increasing $\Gamma$, the zigzag moment evolves from easy-axis ($\text{ZZ}_\parallel$) or coplanar directions ($\text{ZZ}_\perp$) to $\left< \bar{1}11 \right>$ directions ($\text{ZZ}_{ \bar{1}11}$) .
	The system is more frustrated for anti-ferromagnetic $\Gamma^{\prime}$ (c). The zigzag phase closely competes with other orders and a broad IP region.
	$\text{ZZ}_1$ and $\text{ZZ}_2$ distinguish different zigzag configurations.
	Panels (a) and (b) are learned with a symmetric $4 \times 4$ cluster, and panel (c) uses a $6 \times 6$ cluster.
	Colors reflect the value of Fiedler entries at each $(\theta, J)$ points, whose histograms are provided in Appendix~\ref{app:learning}.
	}
 \label{fig:pd_J3}
 \end{figure*}
 
 \begin{figure}[t]
  \centering
  \includegraphics[width=0.4\textwidth]{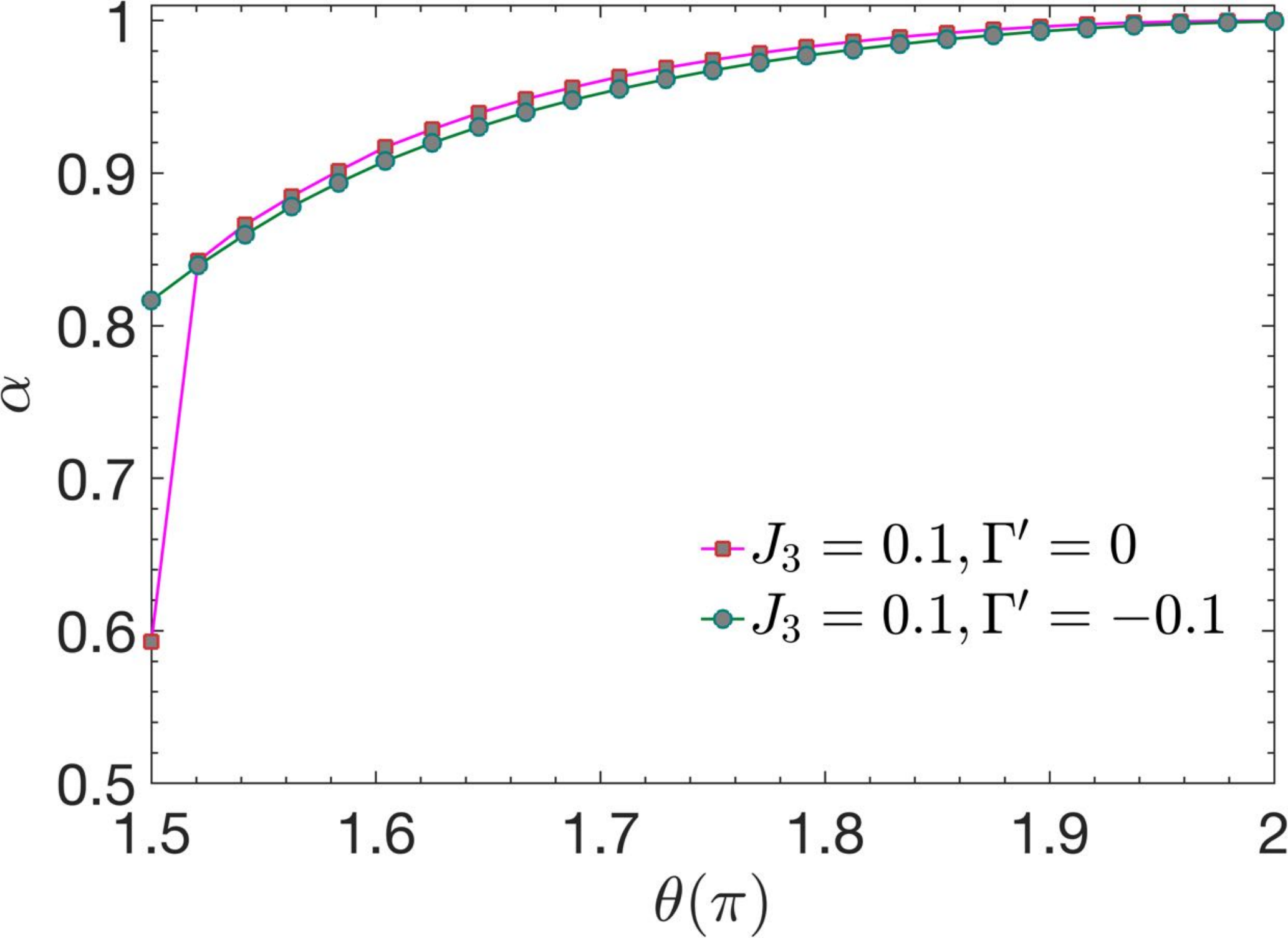}
  \caption{Evolution of the zigzag moment ($\mb{m}_{\rm ZZ}$) along the $J = -0.1$ line, for $ J_3 = 0.1, \ \Gamma^{\prime} = 0, -0.1$, at $T = 10^{-3}$. Spins prefer easy-axes (easy-planes) for the case of $\Gamma^{\prime} = 0$ ($\Gamma^{\prime} = -0.1$) at small $\Gamma$, but evolves towards  $\mb{n} \in \left<\bar{1}11\right>$ with increasing $\Gamma$. $\alpha = \corr{\abs{\mb{m}_{\rm ZZ} \cdot \mb{n}}}$ measures the projection of the magnetic moment on directions of $\left< \bar{1}11 \right>$, and $\corr{.}$ denotes an ensemble average.}
  \label{fig:alpha}
\end{figure}
 
\section{Effects of $\Gamma^{\prime}$ and $J_3$ term}\label{sec:j3gmp}
In modeling of Kitaev materials, the inclusion of the off-diagonal $\Gamma^{\prime}$ and third-neighbor Heisenberg $J_3$ exchange terms can have a phenomenological motivation or a microscopic origin, as discussed in Section~\ref{sec:model}.
In this section, we investigate their effects on the $J$-$K$-$\Gamma$ phase diagram.
 
\subsection{Finite $\Gamma^{\prime}$}\label{sec:gmp}
To disentangle their effects, we first study the case of $J_3 = 0$ and  a finite $\Gamma^{\prime}$.
The major consequence of adding a small ferromagnetic $\Gamma^{\prime} = -0.1$ is that the zigzag order in the $J$-$K$-$\Gamma$ phase diagram expands significantly and prevails over the phase diagram, as plotted in Figure~\ref{fig:pd_gmp} (a).
In addition, a type of $120^{\circ}$ order~\cite{Rau14,Rusnacko19} or anti-ferromagnetic $S_3$ order according to its order parameter structure~\cite{Liu21}, which originally lives in the $K>0, \Gamma > 0$ region, is induced in the corner of large $\Gamma$ and small $J$.
These results are consistent with the observations in Ref.~\onlinecite{Rusnacko19} for the quantum spin-$1/2$ model.

Our machine finds more intricate physics for the $\Gamma^{\prime} = 0.1$ case.
As we show in Figure~\ref{fig:pd_gmp} (b), there are three stable magnetic phases.
A ferromagnet and the nested ZZ-ST magnet dominate the parameter regions of small and large $\Gamma$, respectively, while the large $\Gamma$ and small $J$ limit accommodates an anti-ferromagnet.
These phases are separated by broad crossovers. In particular, as shown in Figure~\ref{fig:mags_gmp} along the $J = 0$ line, in the regime between the nested and anti-ferromagnatic phase, no strong ordering is observed even down to the low-temperature $T = 10^{-4}$.
These regimes are hence also considered incommensurate or correlated paramagnetic (IP), similar as in the previous section for the $\Gamma^{\prime} = 0$ case.

\subsection{Finite $J_3$ and $\Gamma^{\prime}$}\label{sec:J3}

We now compile all the exchange interactions together.
As shown in Figure~\ref{fig:pd_J3} (a) and (b), the anti-ferromagetic $J_3$ exchange term strongly favors the zigzag order regardless of a vanishing or negative $\Gamma^{\prime}$, resulting in a simple topology to the phase diagram.
As measured in Figure~\ref{fig:alpha}, the zigzag moment prefers the directions of easy axes for $\Gamma^{\prime} = 0$ and easy planes for $\Gamma^{\prime} = -0.1$ at small $\Gamma = 0$, and evolves towards $\left<\bar{1}11\right>$ directions upon increasing $\Gamma$.
An exception in the phase diagram is a small incommensurate or disordered area in the top left corner.
This regime may be a remnant of a spiral order in the $O(3)$-symmetric $J_1$-$J_3$ honeycomb Heisenberg model~\cite{Rastelli80, Fouet01}.
It is present only in a narrow window around $J_3/J \sim -\frac{1}{3}$ along the $\Gamma = 0$ line and becomes a trivial ferromagnet at larger $\abs{J}$.

The combination of a positive $J_3$ and positive $\Gamma^{\prime}$ leads to a more complex topology, as shown in Figure~\ref{fig:pd_J3} (c).
While the zigzag phase still dominates the phase diagram, the ferromagnetic and the $S_3 \times Z_3$ phase, relevant for the pure $J$-$K$-$\Gamma$ model (Figure~\ref{fig:pd_0_0}), and the anti-ferromagnetic phase in the vanishing $J_3$ but positive $\Gamma^{\prime}$ case (Figure~\ref{fig:pd_gmp}), reappear.
The nested ZZ-ST order, which occupied a considerable area in the $J_3 = 0$ phase diagrams, is now taken over by an IP regime and a zigzag order.
Clearly, a positive $\Gamma^{\prime}$ competes with $J_3$ and adds frustration.

\begin{table*}[t]
\centering
\renewcommand{\arraystretch}{1.2}
	\begin{tabular*}{\textwidth}{c@{\extracolsep{\fill}} ccccc}
	\toprule
	\hline
	 & $J$ & $K$ & $\Gamma$ & $\Gamma^{\prime}$ & $J_3$ \\
	\midrule
	$\alpha$-${\rm RuCl}_3$\cite{Winter16} & $-0.97$ & $-8.21$ & $4.16$ & $-0.93$ & \\
	$\alpha$-${\rm RuCl}_3$\cite{Kim16} & $-1.67$ & $-6.67$ & $6.6$ & $-0.87$ & $2.8$\\
	$\alpha$-${\rm RuCl}_3$\cite{Winter17} & $-0.5$ & $-5.0$ & $2.5$ & & $0.5$\\
	$\alpha$-${\rm RuCl}_3$\cite{Maksimov20} & $[-4.1,-2.1]$ & $[-11,-3.8]$ & $[3.9, 5.0]$ & [2.2,3.1] & $[2.2,3.1]$ \\
		${\rm Na}_2{\rm Co}_2 {\rm TeO}_6$\cite{Songvilay20} & $-0.1(8)$ & $-9.0(5)$ & $1.8(5)$ & $0.3(3)$ & $0.9(3)$\\
	${\rm Na}_3{\rm Co}_2 {\rm SbO}_6$\cite{Songvilay20} & $-2.0(5)$ & $-9.0(10)$ & $0.3(3)$ & $-0.8(2)$ & $0.8(2)$ \\
	\hline
  \bottomrule
\end{tabular*}
\caption{A selection of representative microscopic interactions (in meV) proposed for three Kitaev materials. A more complete collection of models suggested for $\alpha$-${\rm RuCl}_3$ can be found in Refs.~\onlinecite{Laurell20} and~\onlinecite{Maksimov20}.}
\label{tab:materials}
\end{table*}

\section{Implication to materials}~\label{sec:materials}
We now apply the machine-learned phase diagrams to the representative parameter sets proposed for the compounds mentioned in Section~\ref{sec:model} and reproduced in  Table~\ref{tab:materials}.

Following the parameters given in Ref.~\onlinecite{Songvilay20} based on inelastic neutron scattering (INS), the two cobaltate systems ${\rm Na}_2{\rm Co}_2 {\rm TeO}_6$ and ${\rm Na}_3{\rm Co}_2 {\rm SbO}_6$ both have a dominating ferromagnetic Kitaev exchange and a small anti-ferromagnetic $J_3 \sim 0.1 \abs{K}$.
They fall relatively deeper in the zigzag phase shown respectively in the phase diagrams of Figure~\ref{fig:pd_J3} (a) and Figure~\ref{fig:pd_J3} (b), in agreeement with the experimental result of Ref.~\onlinecite{Songvilay20}.

The compound $\alpha$-${\rm RuCl}_3$ also resides inside the zigzag phase, provided that the $\Gamma^{\prime}$ term is negligible, as suggested by the {\it ab initio} calculation and the INS fit of Ref.~\onlinecite{Winter17}, or is negative, as suggested by the DFT calculations of Refs.~\onlinecite{Kim16, Winter16}.
Nevertheless, if $\Gamma^{\prime}$ is antiferromagnetic, as was recently advocated in Ref.~\onlinecite{Maksimov20}, it falls into the far more complex phase diagram Figure~\ref{fig:pd_J3}~(c).
Consistent with the linear spin-wave analysis of Ref.~\onlinecite{Maksimov20}, the zigzag-like magnet $\alpha$-${\rm RuCl}_3$ will be then adjacent to an incommensurate or disordered regime.

However, it is interesting that, as indicated by Figure~\ref{fig:pd_0_0}, the projection of $\alpha$-${\rm RuCl}_3$ on the $J$-$K$-$\Gamma$ subspace lies close to the frontier of several phases.
Consider the commonly suggested range for the three major exchange interactions of this compound, $\Gamma \sim 0.5\text{--} 1 \abs{K}$ ($\theta \sim 1.65\text{--}1.75 \pi$), $J \sim -0.1\abs{K}$~\cite{Winter16, Ran17, Kim16, Winter17, Laurell20, Maksimov20}.
The relevant area in the $J$-$K$-$\Gamma$ phase diagram Figure~\ref{fig:pd_0_0} encloses, or is close to the boundary of, the $S_3 \times Z_3$, the nested ZZ-ST, a ferromagnetic phase and a broad paramagnetic regime.
These phases may compete with the zigzag order stabilized by a finite $\Gamma^{\prime}$ and/or $J_3$ term, in particular if $\Gamma^{\prime}$ is anti-ferromagnetic.

\section{Summary}\label{sec:summary}
Kitaev materials are promising hosts of exotic phases and unconventionally ordered states of matter. Identifying the nature of those phases and constructing the associated phase diagrams is a daunting task.
In this work we have utilized an interpretable and unsupervised machine-learning method, the tensorial kernel support vector machine (TK-SVM), to learn the phase diagram of a generalized Heisenberg-Kitaev-$\Gamma$ model on a  honeycomb lattice.

Based on data from classical Monte Carlo simulations on large lattice size, the machine successfully reproduces the known magnetic orders as well as the incommensurate or paramagnetic-like regimes reported in the previous quantum and classical studies. 
It also goes further by detecting new phases in the parameter regions relevant for the compounds $\alpha$-${\rm RuCl}_3$, ${\rm Na}_2{\rm Co}_2 {\rm TeO}_6$ and ${\rm Na}_3{\rm Co}_2 {\rm SbO}_6$, including a nested zigzag-stripy phase and showing the extension of the modulated $S_3 \times Z_3$ phase under finite Heisenberg interactions (Section~\ref{sec:jkgm}).
In particular, the machine-learned phase diagrams suggest that, in the $J$-$K$-$\Gamma$ subspace, the actively studied compound $\alpha$-${\rm RuCl}_3$ is situated near the boundary of several competing phases, including a simple ferromagnet, the more involved  $S_3 \times Z_3$ and nested zigzag-stripy magnets, and a possibly correlated paramagnet.
The inclusion of further couplings such as $\Gamma^{\prime}$ and $J_3$ terms stabilizes zigzag order as known in the literature.
However, if the $\Gamma^{\prime}$ exchange in this material is anti-ferromagnetic and sufficiently strong to compete with $J_3$, as recently put forward in Ref.~\onlinecite{Maksimov20}, the proposed parameter set will be adjacent to an incommensurate or correlated paramagnetic regime which may originate from the competition of the magnetic orders indicated above (Section~\ref{sec:j3gmp}).

To simulate large system sizes that unbiasedly accommodate different competing orders, we have treated spins as classical $O(3)$ vectors, corresponding to the large-$S$ limit of quantum spins.
The fate of the novel orders identified by our machine against quantum fluctuations needs to be examined by future studies.
Nevertheless, as strong symmetry-broken orders dominate, our phase diagrams can act as a useful reference for future quantum simulations.
Moreover, by recognizing the unconventional orders and indicating the paramagnetic-like regimes, our phase diagrams may also guide the understanding of existing Kitaev materials and the search for new materials.

We hope our study also stimulates future machine-learning applications in Kitaev materials.
While such systems are motivated by material realizations~\cite{Jackeli09, Chaloupka10} of the Kitaev spin liquid~\cite{Kitaev06}, the presence of various non-Kitaev interactions appears to modify the physics expected from a pure Kitaev model considerably.
Those interactions span a multi-dimensional parameter space, with accumulated evidence of complicated and rich phase diagrams.
Machine learning techniques are designed to discover important information and structure from high-dimensional complex data and can provide new toolboxes for investigating the physics of Kitaev materials and general frustrated magnets.

\begin{acknowledgements}
NR, KL, MM, and LP acknowledge support from FP7/ERC Consolidator Grant QSIMCORR, No. 771891, and the Deutsche Forschungsgemeinschaft (DFG, German Research Foundation) under Germany's Excellence Strategy -- EXC-2111 -- 390814868.
Our simulations make use of the $\nu$-SVM formulation~\cite{Scholkopf00}, the LIBSVM library~\cite{Chang01, Chang11}, and the ALPS\-Core library~\cite{Gaenko17}.
The TK-SVM library has been made openly available with documentation and examples~\cite{Jonas}.
The data used in this work are available upon request.
\end{acknowledgements}

\begin{appendix}
 \begin{figure*}
 \includegraphics[width=0.88\textwidth]{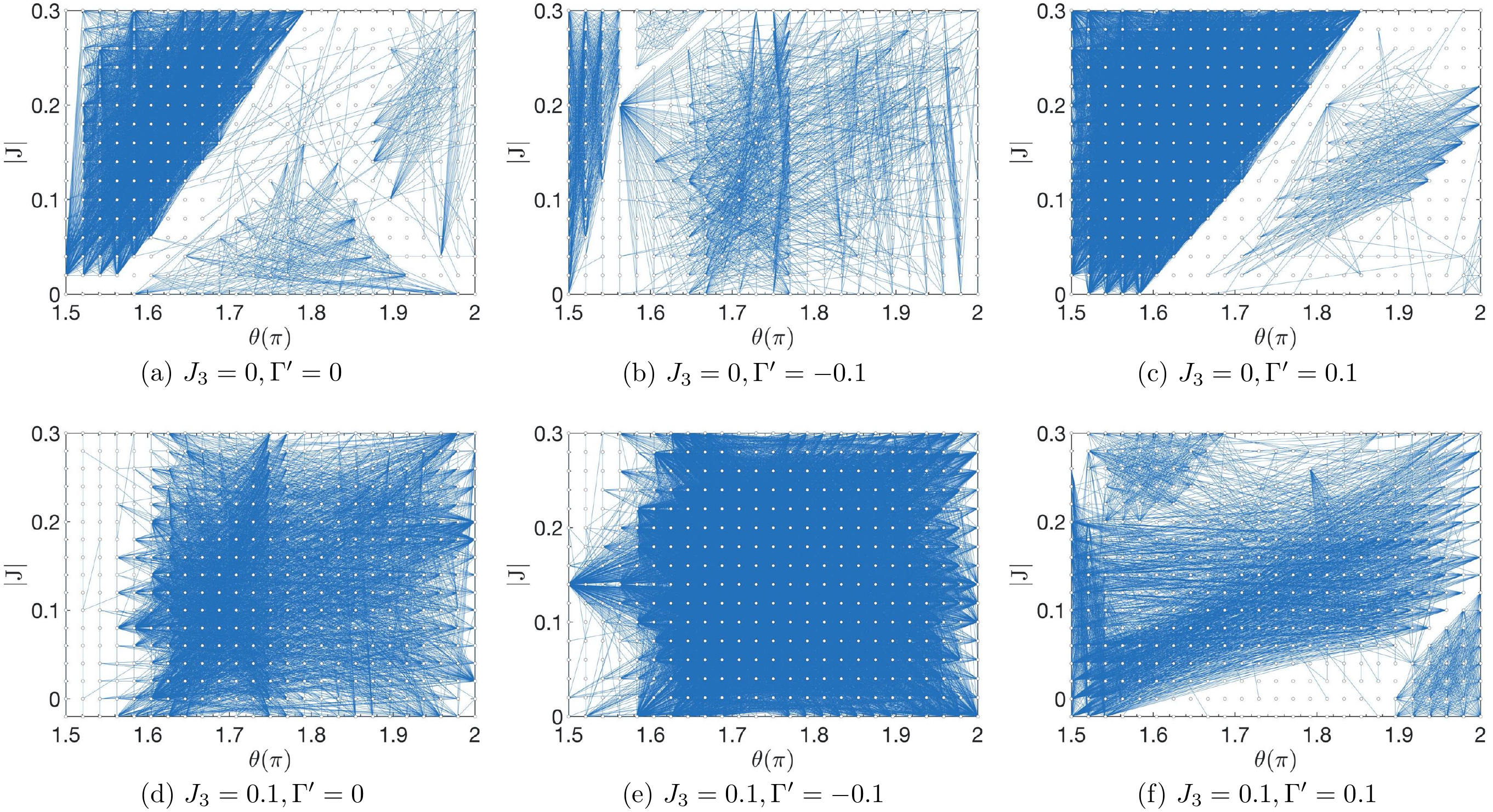}
 \caption{Graphs associated with the phase diagrams discussed in the main text. Each vertex (white circle) represents a $(\theta, J)$-point with fixed $J_3$ and $\Gamma^{\prime}$, from which training samples are collected. The edge (blue line) connecting two vertices is determined by the learned bias parameter  $\rho$. Here $\rho_c = 10^4$ is imposed in the weight function Eq.~\eqref{eq:weight}.
 Each graph contains $400$ vertices and $79,800$ edges. Edge weights are suppressed in the figure for visualization purposes. }
 \label{fig:graphs}
 \end{figure*}

\section{TK-SVM}\label{app:tksvm}
Here we briefly review the TK-SVM method and refer to our previous work in Refs.~\onlinecite{Greitemann19, Liu19, Greitemann19b, Jonas_thesis} for a comprehensive introduction.

\subsection{Decision function}
In the language of TK-SVM, a phase classification problem is solved by learning a binary decision function 
\begin{align}\label{eq:d(x)}
    d(\mb{x}) = \sum_{\mu\nu} C_{\mu\nu} \phi_\mu(\mb{x}) \phi_\nu(\mb{x}) - \rho.
\end{align}
Here $\mb{x} = \{S_{i,a}\}$ represents a real-space snapshot of the system and serves as a training sample, with $i$ and $a$ respectively labeling the lattice index and component of a spin.
$\bds{\phi}(\mb{x})$ is a feature vector mapping $\mb{x}$ into degree-$n$ monomials,
\begin{align}\label{eq:phi}
   \phi^{(n)}: \mb{x} \rightarrow \ \bds{\phi}(\mb{x}) = \{\phi_{\mu}\} = \{\corr{S_{\alpha_1}^{a_1} S_{\alpha_2}^{a_2} \dots S_{\alpha_n}^{a_n}}_{\rm cl} \},
\end{align}
where $\corr{\dots}_{\rm cl}$ is a lattice average over finite clusters of $r$ spins, $\mu = \{\alpha_n, a_n\}$ denotes a collective index, $\alpha_n$ labels spins in a cluster, and the degree $n$ defines the rank of a TK-SVM kernel.
The map $\phi^{(n)}$ is based on the observation that a symmetry-breaking order parameter or a local constraint for rotor degrees of freedom can be in general be represented by finite-rank tensors or polynomials~\cite{Liu16,Nissinen16,Michel01}.
With this map, the decision function probes both linear and higher-order correlators, including magnetic order, multipolar order and ground-state constraints~\cite{Greitemann19,Liu19,Greitemann19b}.
Moreover, this map can be combined with other machine-learning architectures, such as a principal component analysis (PCA). However, as elaborated in the thesis of J. Greitemann~\cite{Jonas_thesis}, it was found that TK-SVM has in general better performance and interpretability than TK-PCA. In a recent paper Ref.~\onlinecite{Miles21}, a nonlinear feature map with similar spirit was employed in a novel architecture of convolutional neural networks.

The coefficient matrix $C_{\mu\nu} $ in the decision function identifies important correlators that distinguish two data sets, from which order parameters can be extracted.
It is defined as a weighted sum of support vectors,
\begin{align}\label{eq:C_munu}
    C_{\mu \nu} = \sum_k \lambda_k \phi_\mu(\mathbf{x}^{(k)}) \phi_\nu(\mathbf{x}^{(k)}),
\end{align}
where $\lambda_k$ is a Lagrange multiplier and represents the weight of the $k$-th support vector.

The bias $\rho$ in Eq.~\eqref{eq:d(x)} is a normalization parameter in a standard SVM, but in TK-SVM it is endowed a physical implication to detect phase transitions and crossovers, or the absence thereof\mbox{~\cite{Liu16, Greitemann19b}}.
For two sample sets $p$ and $q$, it behaves as
\begin{align}\label{eq:rho_rules}
    \abs{\rho_{pq}} \begin{cases}
        \gg 1 & \textup{${p, q}$ in the same phase}, \\
        \lesssim 1 & \textup{${p, q}$ in different phases}.
    \end{cases}
\end{align}
Although the sign of $\rho$ can indicate which data set is more disordered, its absolute value suffices to construct a phase diagram; cf. Ref.~\onlinecite{Greitemann19b} for a comprehensive discussion.

The above binary classification is straightforwardly extended to a multi-classification problem over $M > 1$ sets.
SVM will then learn $M(M-1)/2$ binary decision functions, comprising binary classifiers for each pair of sample sets~\cite{HsuLin02}.

\begin{figure*}
 \includegraphics[width=0.88\textwidth]{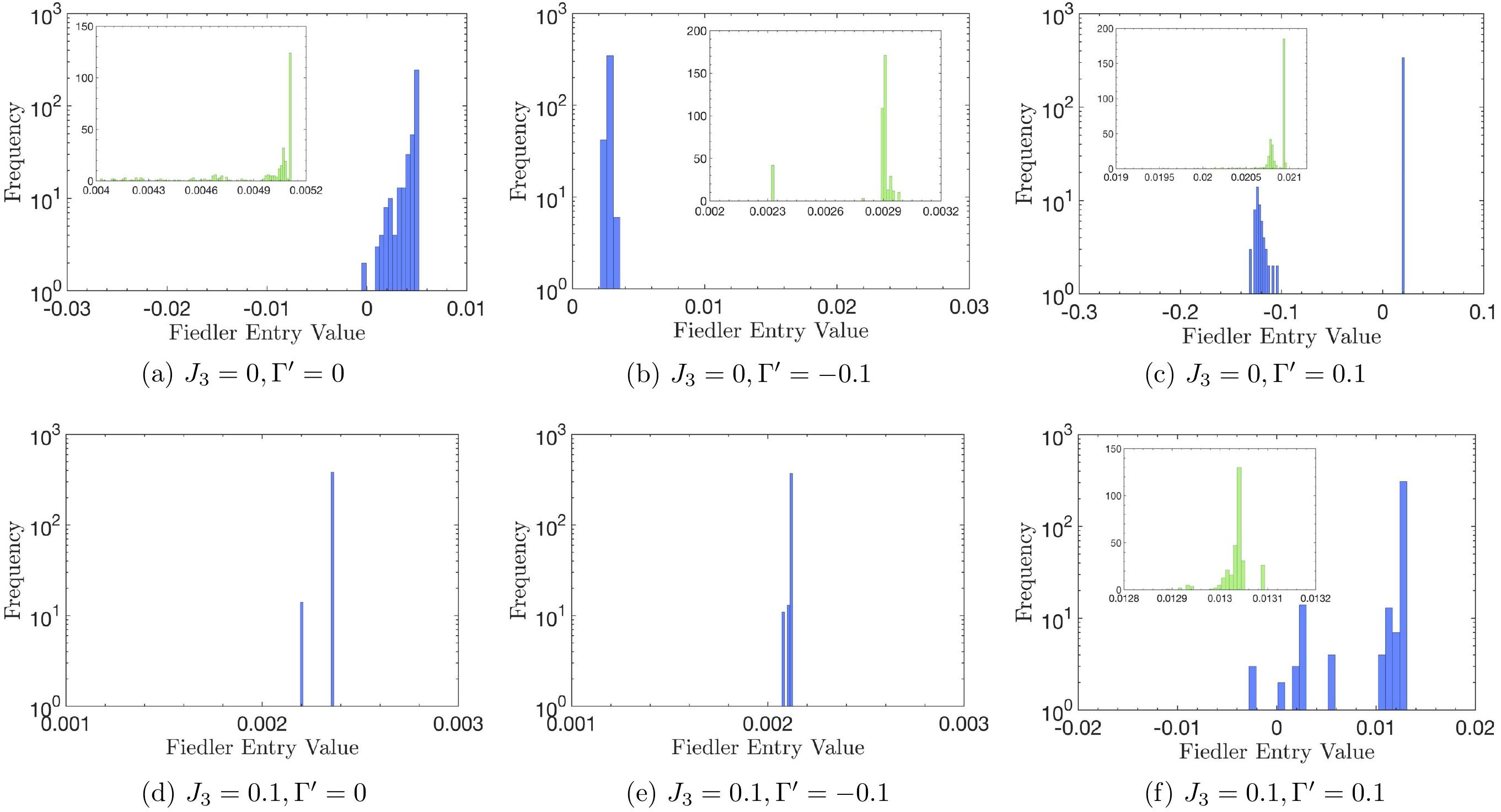}
 \caption{Histograms for the dominating Fiedler entries of the six graph partitioning problems. The main panels have a logarithmic scale on the vertical axis because the distribution spans several orders of magnitude. The insets show the main part of the distribution on a linear scale for easier comparison.}
 \label{fig:hist_gp_J3}
 \end{figure*}
 
\subsection{Graph partitioning}
Although the standard SVM is a supervised machine-learning method~\cite{BookVapnik}, the supervision can be skipped in the TK-SVM framework thanks to multi-classification and graph partitioning.

A graph $G = (V, E)$ can be viewed as a pair of a vertex set $V$ and an edge set $E$ connecting vertices in $V$.
Each vertex represents a phase point in the physical parameter space where we collect training data.
For the $J$-$K$-$\Gamma$ phase diagram with fixed $\Gamma^{\prime}$ and $J_3$ couplings, these vertices are specified by the value of $\{\theta, J\}$.
We work with weighted graphs. Namely, the edge linking two vertices $v_p, v_q \in V$ has a weight $w_{pq} \in [0, 1)$.
Intuitively, if $v_p, v_q$ are in the same phase, they will be connected with a large $w_{pq}$; otherwise $w_{pq} \simeq 0$.

The weight of an edge is determined by the bias parameter, according to its behavior given in Eq~\eqref{eq:rho_rules}.
The choice of the weighting function turns out not to be crucial.  We adopt the normal Lorentzian weight distribution,
\begin{align}\label{eq:weight}
  w(\rho) &= 1 - \frac{\rho_c^2}{(|\rho|-1)^2+\rho_c^2} \in [0,1),
\end{align}
where $\rho_c$ is a hyperparameter setting a characteristic scale to quantify ``$\gg 1$''.
The choice of $\rho_c$ is uncritical and does not rely on fine tuning, as edges connecting vertices in the same phase have always larger weights than those crossing a phase boundary.
In practical use, we vary $\rho_c$ over several orders of magnitude to ensure the results are robust.
In this work, $\rho_c = 10^2, 10^3, 10^4$ are considered, and $\rho_c = 10^4$ is chosen to construct the graphs in Fig.~\ref{fig:graphs} and the  subsequent phase diagrams.
We refer to our previous works Refs.~\onlinecite{Greitemann19b} and~\onlinecite{Rao21} for comparisons of results using different $\rho_c$'s.

A graph of $M$ vertices and $M(M-1)/2$ edges can be represented by a $M\times M$ Laplacian matrix 
\begin{align}
	\hat{L} = \hat{D} - \hat{A}.
\end{align}
Here, $\hat{A}$ is a symmetric off-diagonal adjacency matrix with $A_{pq} = w_{pq}$ hosting the weights of the edges.
$\hat{D}$ is a diagonal degree matrix where $D_{pp} = \sum_{p \neq q} w_{pq}$ denotes the degree of vertices.

We then utilize Fiedler's theory of spectral clustering to partition the graph $G$~\cite{Fiedler73, Fiedler75}, which is achieved by solving for the eigenvalues and eigenvectors of $\hat{L}$.
The second smallest eigenvalue $\lambda_2$ reflects the algebraic connectivity of the graph, while the respective eigenvector $\mathbf{f}_2$ is known as the Fiedler vector.
Entries of $\mathbf{f}_2$ have a one-to-one correspondence with vertices of the graph.
Vertices (the physical parameter points) in the same subgraph will be assigned nearly identical Fiedler entries, while those in different subgraphs will be assigned contrasting values.
The Fiedler vector can thereby effectively act as a phase diagram.

\section{Setup of the sampling and learning}\label{app:learning}
The parameters specified in Section~\ref{sec:model} lead to six individual problems, depending on the value of $\Gamma^{\prime}$ and $J_3$.
For fixed $\Gamma^{\prime}$ and $J_3$, $400$ phase points are simulated in the $(J, \theta)$ subspace and $500$ configurations are sampled at each point.
These phase points distribute uniformly in the parameter range $J \in [-0.3, 0]$ and $\theta \in [1.5 \pi, 2 \pi]$, with $\Delta J = 0.02$, $\Delta \theta = \frac{1}{48} \pi$.
Such a protocol of sampling does not reflect a particular strategy but just represents a natural choice when exploring unknown phase diagrams. 

We perform a TK-SVM multi-classification analysis on the sampled data with different clusters and ranks in the map $ \phi^{(n)}$ in Eq.\eqref{eq:phi}.
Each learning problem comprises $79,800$ binary decision functions, and a graph with $400$ vertices and $79,800$ edges is constructed from the learned $\rho$ parameters, as visualized  in Figure~\ref{fig:graphs}.
In all six cases, the phase diagrams can be mapped out with just rank-$1$ TK-SVMs, while a universal choice of the cluster is simply choosing a symmetric cluster with $m \times m$ honeycomb unit cells (see Figure~\ref{fig:lattice}).
We confirm the consistency of a phase diagram by checking the results against the ones found when using larger clusters with $m = 4,\, 6, \, 12$ ($32, \, 72, \, 288$ spins). 

The partitioning of these graphs leads to Fiedler vectors, which reveal the topology of the phase diagrams, and are color plotted in the main text.
Figure~\ref{fig:hist_gp_J3} shows the histograms of the Fiedler vector entires.
The pronounced peaks identify well-separated phases, and the flat regions indicate disordered regimes and crossovers between the phases.

After having determined the topology of the phase diagram, the coefficient matrix $C_{\mu\nu} $  is analyzed in order to extract the order parameter for distinct phases.
In cases where no magnetic order is detected, we additionally perform a rank-$2$ TK-SVM analysis and identify a phase as a spin liquid if there is a stable ground-state constraint or as a correlated paramagnet or incommensurate phase if such a constraint is absent.  
The learned order parameters as well as the phase diagrams are validated by additional Monte Carlo simulations in Sections~\ref{sec:jkgm} and~\ref{sec:j3gmp}.

\section{Simulation Details}\label{app:simulation}
We use parallel tempering (PT) in combination with the heat bath and over relaxation algorithms to equilibrate the system.
The distribution of temperatures are carefully chosen to ensure efficient iterations between different temperatures~\cite{Katzgraber06}.  
In the training stage, $N_T = 64$ logarithmically equidistant temperatures between $T = 0.5 \times 10^{-3}$ and $10$ are used for the majority of the $(\theta, J)$ points, while $N_T = 128$ temperatures are needed for a small subset of special parameter points.
In the testing stage, $N_T = 128$ temperatures in $[0.75 \times 10^{-4}, 10]$ are sufficient for most points measured in Figs.~\ref{fig:mags}, \ref{fig:mag_nest} and \ref{fig:mags_gmp}, but $N_T = 256$ temperatures are required for a few points. 

We typically run $8 \times 10^6$ Monte Carlo (MC) sweeps for simulations using $N_T = 64$ and $1.6\times 10^7$ sweeps for those requiring more temperatures.
Half of the total sweeps are considered as thermalization.
In the training stage, $500$ samples are collected from the second half of simulations. Namely, the sampling interval is $800$ or $1,600$ MC sweeps.
In the testing stage, new Monte Carlo simulations are performed to measure the learned order parameters (Figs.~\ref{fig:mags}, \ref{fig:mag_nest} and \ref{fig:mags_gmp}), and several independent simulations are compared to confirm the ergodicity and good thermalization of our simulations.
Alternatively, one could also measure the learned decision functions (without interpretation) as in applications of a standard SVM. 
As we established in Refs.~\onlinecite{Greitemann19, Liu19}, the TK-SVM decision function is essentially an encoder of physical order parameters.
However, the goal of machine-learned phase diagrams is to not only to assign each phase a different label but also to find suitable characterizations of the phases.
Direct measurements of physical quantities are hence preferred if interpretability is guaranteed.

\end{appendix}
\bibliography{jkgm}

\end{document}